\documentclass[aps,pra, groupedaddress, floatfix, twocolumn]{revtex4-2}
\usepackage{amsmath}
\usepackage{mathrsfs}
\usepackage{amsfonts}
\usepackage{wrapfig}
\usepackage{stmaryrd}
\usepackage{xcolor}

\usepackage{graphicx}
\usepackage{float}
\usepackage{subfig}
\usepackage{subcaption}
\usepackage[hidelinks]{hyperref}

\begin{document}

\title{Collective emission of atomic nanorings around an optical nanofiber}
\author{J. Jim\'enez-Jaimes$^1$, S. Nic Chormaic$^2$, E. Brion$^1$}
\affiliation{$^1$Laboratoire Collisions Agrégats Réactivité, UMR 5589, FeRMI, Université de
Toulouse, CNRS, 118 Route de Narbonne, F-31062 Toulouse, France,\\
$^2$OIST Graduate University, Onna-son, Okinawa 904-0495, Japan }

\begin{abstract}
We theoretically investigate the collective emission of one and two  circular arrays of two-level atoms surrounding an optical nanofiber. 
We show that the radiation eigenmodes of a single ring selectively couple to specific guided modes of the fiber, according to their symmetry, and study how the physical parameters of the system control their nature.  In particular, we identify situations in which the emission toward radiation modes is highly suppressed with respect to fiber-guided modes while the lifetime of the atomic excitation is enhanced. 
We further address the case of two identical nanorings positioned at a distance from each other along the nanofiber. By contrast to free-space configurations,  the rings can  exchange excitations even at large separation through nanofiber guided modes resulting in 
enhanced sub- and supperradiance with respect to the single-ring case. 
Our findings suggest that the ring configuration is promising for the implementation of efficient and versatile light-matter nanofiber-based interfaces and the achievement of waveguide quantum electrodynamics.

\end{abstract}

\maketitle


\section{Introduction}

Light-matter interactions are at the core of most proposals and implementations in quantum technologies. Ongoing progresses in nanofabrication make it possible to design nanowaveguides that offer both strong light confinement and tailored optical guided modes. Tapered subwavelength optical fibers, also called optical nanofibers (ONFs), are among the most versatile and promising light-matter interfaces \cite{LBV24}. Relatively easy to fabricate in the laboratory, ONFs possess a great wealth of guided modes \cite{FPT12, HFB15}  with an intense evanescent component \cite{LLH04,LBT17}. Interaction with these modes is instrumental to detect and trap quantum emitters such as atoms, molecules, quantum dots or NV centers in diamond around the ONF  \cite{Kumar_2015,PhysRevA.96.043859, NGN16, YLM12, L14, PSW16, WVM07, LBH04, FYT07, VRS10}. The generation, lifetime,  interactions, and trapping of highly excited  Rydberg atoms in the vicinity of a silica ONF have also been investigated,  both theoretically and experimentally \cite{SZL19, SLR20, SLR23, PhysRevResearch.2.012038, VBR23}. Due to the 
interactions ONFs provide  between a single guided mode and a large number of surrounding atoms, nanofiber-based platforms also pave the way to the implementation of waveguide quantum electrodynamics (WQED)  \cite{SPI23}  and the observation of collective phenomena such as super-, sub- or selective radiance \cite{AMA17}. Superradiance was  already demonstrated in a system of two atomic clouds 0.3 mm apart from each other and coupled via the evanescent field of an optical nanofiber  \cite{SBF17}.  A collective excitation was also imprinted then read out via the DLCZ protocol in an  array of 
$^{133}$ Cs atoms trapped along an ONF \cite{CRC19}. EIT-based slow-light  and light storage were reported both in an atomic cloud overlapping  an ONF \cite{GMN15}  and in a regular array of atoms trapped in an optical lattice realized in the evanescent field around the ONF \cite{SCA15}. Large Bragg reflection was also demonstrated in an atomic array trapped in the evanescent field of an ONF, with both unidirectional and chiral interactions observed \cite{CGC16}.

Subradiant states of ordered atomic arrays have been particularly studied,  theoretically \cite{ SML12, KSP16, JCC18, PSR19, ZM19, FSK21}  and experimentally  \cite{PCP85, DB96, GAK16, RWR20, FGH21}, for their potential applications to quantum memories \cite{AMA17} and transport \cite{NLO19}. The control of their  lifetime was investigated in 1D arrays in free space \cite{AMA17,ZM20, KSK21,FFB23}, near a waveguide \cite{KPL19,SR16, AHC17, KCL19, ZYM20, P20}, as well as in 2D arrays \cite{FJR16, BGA15, BR20}. Because of its unique symmetry properties, the ring geometry was also considered in free-space and shown to potentially lead to enhanced single-photon sensing, transport, storage, and light generation in engineered nanoscale systems \cite{AMA17,JCC18,MPO19,NLO19,CPM20,HPO20, MHR22,CLO23,HPO24,UKV24}.  

Herein we investigate the emission of  
two-level atoms arranged in a regular nanoring configuration 
around an ONF. We show that the different radiation eigenmodes (REMs) of a singly excited ring selectively couple to specific nanofiber guided modes (NFGMs) depending on their symmetry properties and that, by properly  choosing the parameters and state of the system, one can make the ring emit predominantly towards a given NFGM while strongly suppressing losses towards radiation modes (Sec.\ref{SecSingleRing}). We also show that two nanorings can exchange an excitation, even at a large  separation distance, via the NFGMs, and that the REMs of the two-ring system resulting from this coupling exhibit enhanced sub- and superradiance with respect to the single-ring case (Sec.\ref{SecTwoRings}).  Our study is not restricted to the single-mode regime, contrary to most previous works on collective radiance in atomic systems in the vicinity of an ONF \cite{KDN05, KR17}.  Though many other configurations could and will be considered in the future, this work already demonstrates the great potential of interfacing atomic nanorings with an ONF for implementing highly selective and enhanced coupling to NFGMs and therefore entering the realm of WQED.

\section{A single ring}\label{SecSingleRing} 
\begin{figure}
    \captionsetup{width=\linewidth}
    \includegraphics[width = \linewidth]{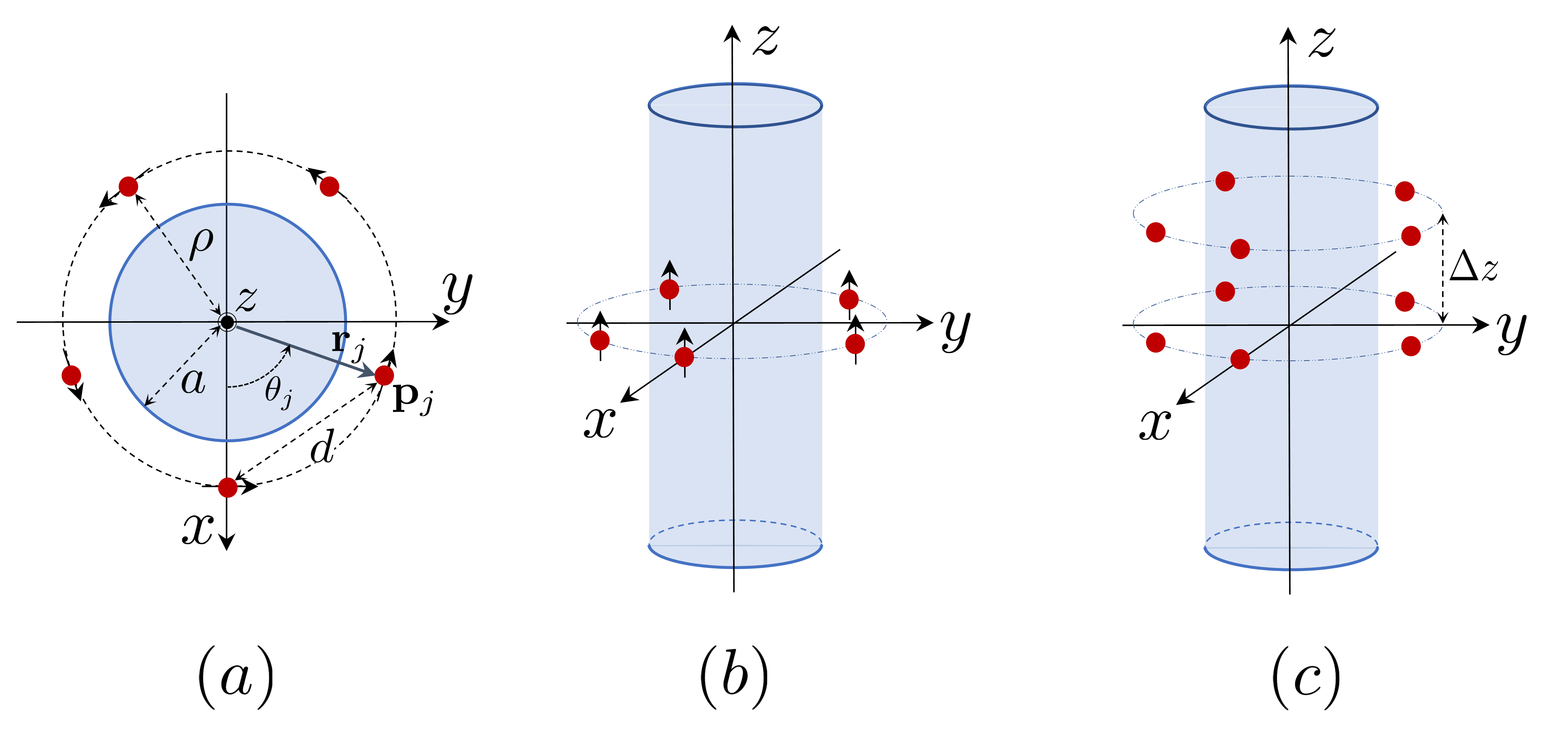}
    \caption{(a,b) Circular array of $N=5$ identical two-level atoms around an ONF of axis $(Oz)$. $\rho$ and $a$ denote the radii of the array and the nanofiber, respectively. The position vector and polar angle of atom $(j)$ are denoted by ${\bf r}_j$ and $\theta_j\equiv \frac{2 \pi \left( j-1 \right)}{N}$, respectively. Atomic dipoles, ${\bf p}_j$, are either (a) orthoradial,  ${\bf p}_j \equiv p \; {\bf e}_\theta\left( \theta_j \right) $, or (b) along the $z$-axis,  ${\bf p}_j \equiv p \; {\bf e}_z $. (c) Two identical nanorings, separated by the distance $\Delta z$.}
    \label{ConfigSystem}
\end{figure}
We start by considering  the case of a single ring comprising $N$ 
two-level atoms surrounding a silica  ONF of optical index $n_\text{fiber}=1.45$  (see  Fig. \ref{ConfigSystem}  for hypotheses and notations). Throughout this article, we fix $N = 5$  ; although this choice is arbitrary it is actually not critical since the typical features obtained in this case are  also observed for $2\leq N \leq 10$, as verified numerically.  When the atomic system is initially singly excited, it evolves according to the effective -- symmetric but nonHermitian -- Hamiltonian \cite{AMA17,MPO19} 
\begin{equation}
H_{\text{eff}} \equiv - \frac{3 \pi  \hbar \gamma_0}{k_0}\sum_{i,j=1}^{N}g_{ij}\;\sigma_{+}^{\left(i\right)}\sigma_{-}^{\left(j\right)}\label{Heffectif}
\end{equation}
where $\gamma_{0}$ is the single-atom spontaneous emission rate in free space,  $g_{ij} \equiv {\bf u}_{i}^{*}\cdot{\bf G}\left({\bf r}_{i},{\bf r}_{j},k_{0}\right)\cdot{\bf u}_{j}$ is the coupling between atoms $i$ and $j$,  ${\bf u}_{i}\equiv\frac{{\bf p}_i }{\left\| {\bf p}_i \right\|}$
is the normalized transition dipole moment of atom $i$, $k_{0}\equiv\frac{\omega_{0}}{c}$ and $\omega_{0}$ are the wavevector and frequency of the atomic transition, respectively, $\sigma_\pm^{\left( i \right)}$  is the $i$th atom raising/lowering operator,  and  ${\bf G}$ is the Maxwell Green's dyadic tensor in the presence of the fiber \citep{SLR20,KKN22,Tai94}. Note that  $\bf{G}$ splits into its free-space, ${\bf G}_0$,  guided, ${\bf G}_g$, and radiation mode contributions, ${\bf G}_r$, and the same decomposition applies to $g_{ij}$'s.  
The eigenvectors and eigenvalues of $H_{\text{eff}}$ are denoted by $\left\{ \left|\Psi_n\right\rangle , - \frac{3 \pi \hbar \gamma_0}{k_0}\lambda_n \right\} $ with the characteristic index $n$ conventionally taking the integer values $n=\lceil - \frac{N-1}{2} \rceil \cdots \lceil  \frac{N-1}{2} \rceil$, where $\lceil\cdot\rceil$ denotes the ceiling value. Since $H_{\text{eff}}$ is not Hermitian, the $\lambda_{n}$'s are \emph{a priori} complex and the $\left|\Psi_n\right\rangle $'s are \emph{a priori} not orthogonal to each other according to the Hermitian scalar product. These states are the ring radiation eigenmodes (REMs) 
 with  associated decay rates $\Gamma_n\equiv - \frac{6 \pi \hbar \gamma_0}{k_0} \text{Im}\lambda_n$. For a regular nanoring, one has \cite{MPO19}  
\begin{align}
\lambda_n & = \frac{1}{N} \sum_{kl} e^{- \text{i} n \frac{2 \pi ( k-l)}{N} }  g_{kl} \label{eqlambda}\\
\left|\Psi_n\right\rangle & = \frac{1}{\sqrt{N}} \sum_{j}  e^{  \text{i} n \frac{2 \pi ( j-1)}{N} } \left|e_j\right\rangle \label{eqPsin}
\end{align}
where $\left|e_j\right\rangle$ denotes the state in which only the $j^{\text{th}}$ atom is excited.

\begin{figure*}
    \includegraphics[width=\textwidth]{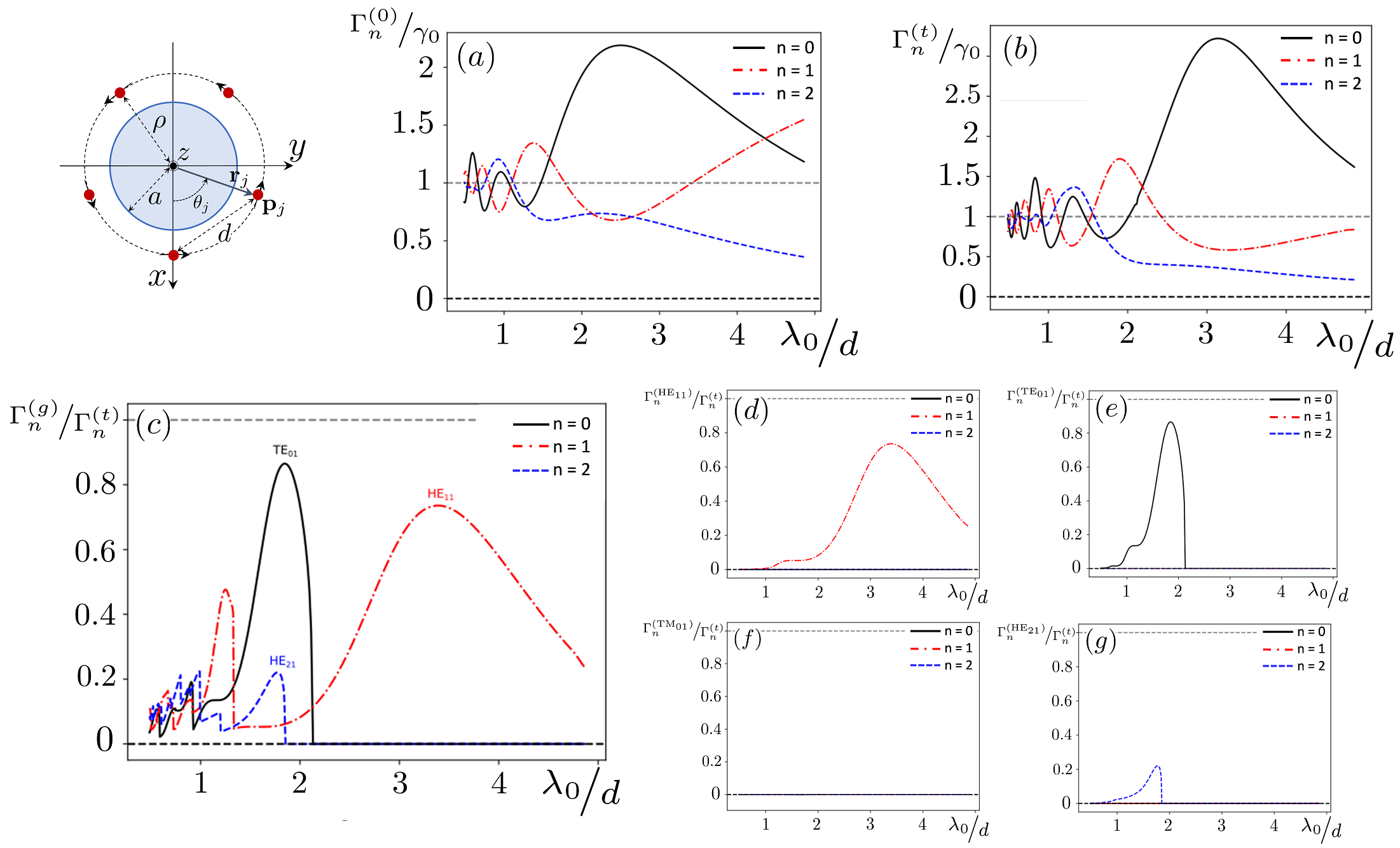}  
    \caption{Nanoring of $N=5$ identical two-level atoms with orthoradial dipoles. Total decay rates of the REM $n=0,1,2$ (a) in free space, $\Gamma_{n}^{\left(0\right)}$,
    and (b) in the presence of the fiber,  $\Gamma_{n}^{\left(t\right)}$, expressed in units of single-atom spontaneous emission rate in free-space $\gamma_0$. 
    (c) Ratio of the decay rate of the REM $n=0,1,2$  towards NFGMs,  $\Gamma_{n}^{\left(g\right)}$, over the total decay rate of the same REM, $\Gamma_{n}^{\left(t\right)}$.      
    Ratio of the decay rate of the REM  $n=0,1,2$ towards NFGMs $\lambda$, $\Gamma_{n}^{\left( \lambda \right)}$, over the total emission rate of the same REM, $\Gamma_{n}^{\left(t\right)}$, for $\lambda=\text{HE}_{11} \; (d),\;\text{TE}_{01}\; (e),\;\text{TM}_{01}\; (f),\;\text{HE}_{21}\;(g)$. In our calculations, we set $\rho=1.1\times a$ whence $d\approx 1.29\times a$, we restrict ourselves to the range $\frac{\pi} {5}a \leq\lambda_{0}\leq2\pi a$, and we take into account $15$ radiation modes and all the guided modes supported by the fiber in the considered frequency range. To provide a visual reference, we have plotted the unit value of the different ratios in a dashed grey line in each subfigure.}
    \label{FigGammaOrtho}   
\end{figure*}

In Fig. \ref{FigGammaOrtho},  we plot the  decay rates of the REMs $n=0,1,2$  as functions of $\frac{\lambda_0}{d}$ -- here  $d=2\rho \sin\left( \frac{\pi}{N} \right)$ is the distance between two neighbouring atoms in the ring -- (a)  in free space, $\Gamma_{n}^{\left(0\right)}$, and (b) in the presence of the fiber, $\Gamma_{n}^{\left(t\right)}$,  in the case of orthoradial dipoles (Fig. \ref{ConfigSystem}a).  
The two plots show similar peak structures, nevertheless, in the close vicinity of the ONF -- here $\rho=1.1 a$,  the strongly collective behavior range is shifted towards higher values of the ratio $\frac{\lambda_0}{d}$, which we attribute to the effect of the ONF optical index, while the superradiant peaks and subradiant dips are more pronounced.

\begin{figure}
\captionsetup{width=\linewidth}
\includegraphics[width = \linewidth]{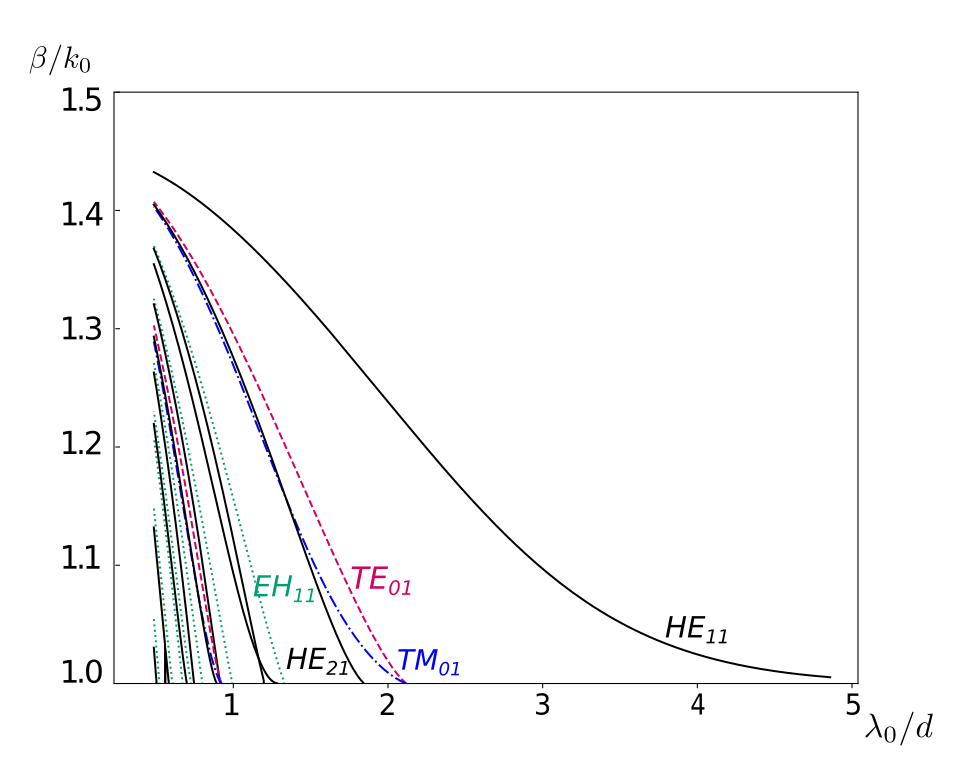}
\caption{Dispersion relation $\frac{\beta}{k_0}=f\left( \frac{\lambda_0}{d} \right)$ of the guided modes of a silica ONF ($n_{\text{fiber}}=1.45$). We only label the first few modes relevant to our study. We set $\frac{\rho}{a}=1.1$  and restrict ourselves to the range $\frac{\pi a}{5} \leq \lambda_0 \leq 2 \pi a$.\label{FigGuidedModes}}
\end{figure}

In Fig. \ref{FigGammaOrtho}c, we plot the ratios  $\frac{\Gamma_{n}^{\left(g\right)}}{\Gamma_{n}^{\left(t\right)}}$ of the decay rates from the  REMs $n=0,1,2$, towards the nanofiber guided modes (NFGMs), $\Gamma_{n}^{\left(g\right)}$, over the total spontaneous emission rates of the same REMs, $\Gamma_{n}^{\left(t\right)}$, as functions of the ratio $\frac{\lambda_0}{d}$. 
Figs. \ref{FigGammaOrtho}d-g 
further show the separate contributions of the different NFGMs which are found to
couple only to specific REMs. 
In a mode-function approach \citep{KBT17}, the coupling between the REM $n$  and a NFGM $\text{K}_{l \geq 0,m}$, with $\text{K} = \text{HE},\;\text{EH},\;\text{TE},\;\text{TM}$, of polarization vector ${\bf e}^{\text{K}}_{lm}$ and azimuthal number 
$pl$, with $l\geq 0$ and  $p=\pm$,
is indeed proportional to 
$\sum_j{e^{\text{i}  \frac{2 \pi}{N} (n+ p l) (j-1)}{\bf u}_{j}\cdot{\bf e}^{\text{K}}_{lm}\left({\bf r}_{j}\right)}$. For orthoradial atomic dipoles,  ${\bf u}_j = {\bf u}_\theta \left( {\bf r}_j \right)$, this term simplifies to 
$\sum_j{e^{\text{i}  \frac{2 \pi}{N} (n+ p l) (j-1)} \left[{\bf e}^{\text{K}}_{lm}\left({\bf r}_{j}\right)\right]_\theta}$.  Since  $\left[{\bf e}^{\text{TM}}_{lm}\left({\bf r}_{j}\right)\right]_\theta = 0$  for all $j=1\cdots N$,  NFGMs $\text{TM}_{0m}$  do not couple to any REM as confirmed by Fig. \ref{FigGammaOrtho}f.  
Moreover, since $\left[{\bf e}^{\text{K=HE,EH,TE}}_{lm}\left({\bf r}_{j}\right)\right]_\theta$ does not depend on the polar angle $\theta$ \citep{KBT17}, the coupling of REMs to NFGMs  $\text{K}_{lm}$  is proportional to 
$\sum_{j=1\cdots N}{e^{\text{i}  \frac{2 \pi}{N} (n+ p l) (j-1)} } = N \times \delta_{n+pl\left\llbracket N \right\rrbracket}$ where 
$\delta_{n+pl\left\llbracket N \right\rrbracket}$  
is nonzero only for $n+pl= kN$ with $k\in\mathbb Z$.  Therefore
NFGMs $\text{TE}_{01},\;\text{HE}_{11},\;\text{HE}_{21}$  -- and more generally  $\text{TE}_{0m},\;\text{HE}_{1m},\;\text{HE}_{2m}$  -- couple to REMs $n=0,\pm1,\pm2$, respectively (Figs.  \ref{FigGammaOrtho}d,e,g). 
In particular,  at  $\frac{\lambda_0}{d}\approx 3.4$, the REM $n=1$  is  slightly subradiant,  $\frac{\Gamma_{n=1}^{(t)}}{\gamma_0}\approx 0.6$  (see Fig. \ref{FigGammaOrtho}b), because the contribution of free-space and radiative modes is strongly suppressed hence maximizing  the probability for the spontaneously emitted photon to be captured by the NFGM  $\text{HE}_{11}$, $\frac{\Gamma_{n=1}^{(\text{HE}_{11})}}{\Gamma_{n=1}^{(t)}}\approx 0.75$.  
Note also that, as $\frac{\lambda_0}{d}$ decreases, higher-order  NFGMs appear and couple to the ring. For instance, HE$_{31}$  the cutoff wavelength of which is around $1.3d$, couples to REMs $n=\pm2$.
This  results in the complex peak structure 
observed in Fig.  \ref{FigGammaOrtho}c.

\begin{figure*}
    \includegraphics[width = 1\textwidth]{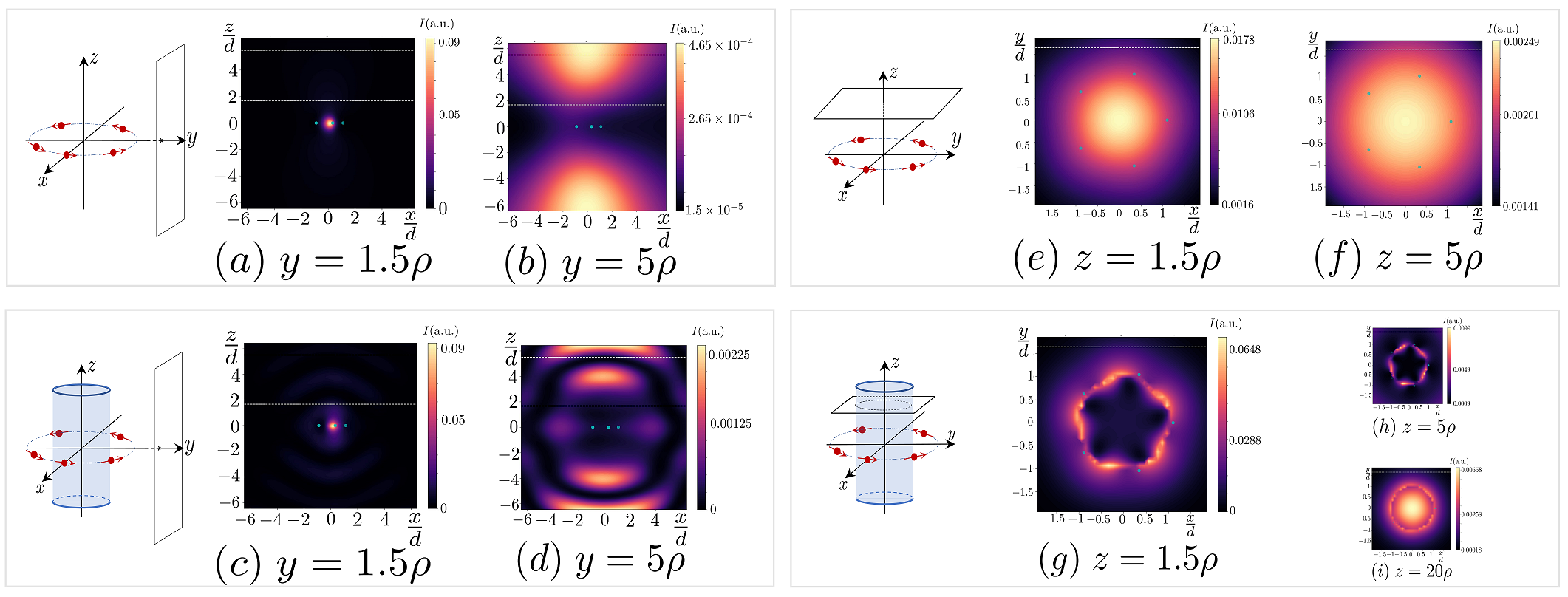}
    \caption{Nanoring of $N=5$ identical two-level atoms with orthoradial dipoles. Radiation patterns of the nanoring prepared in the REM $n=1$ in free space (a,b,e,f) and in the vicinity of an ONF (c,d,g,h,i). The 2D patterns are represented in the $xz$-plane at $y = 1.5 \rho$ (a,c) and $y = 5 \rho$ (b,d), and in the $xy$-plane at $z = 1.5 \rho$ (e,g), $z=5 \rho$ (f,h) and $z=20 \rho$ (i). 
    Intensity is expressed in arbitrary units. 
    The marks of planes $z=1.5\rho$ and $z=5\rho$ appear in dotted white lines in a-d, and the marks of the plane $y=1.5\rho$ appear in dotted white lines in e-i. 
    In our calculations, we set $\rho=1.1\times a$ whence $d\approx 1.29\times a$, we choose $\frac{\lambda_0}{d} \approx 3.4 $ which corresponds to the maximal ratio $\frac{\Gamma_{n=1}^{\left( \text{HE}_{11}\right)}}{\Gamma^{\left(t\right)}_{n=1}}$ (Fig. \ref{FigGammaOrtho}d) and we take into account 15 radiation modes and all the guided modes supported by the fiber at the considered frequency.}
    \label{IntensityOrtho1}    
\end{figure*}

The presence of the ONF also reshapes the radiation patterns. In Fig. \ref{IntensityOrtho1} we plot the intensity radiated by the ring in REM $n=1$ in free space (a,b,e,f) and close to the ONF (c,d,g,h,i) in $xz$-plane at $y=1.5 \rho$ (a,c) and $y=5\rho$ (b,d), and in $xy$-plane at $z=1.5\rho$ (e,g),  $z=5\rho$ (f,h) and $z=20\rho$ (i).  We choose the value $\frac{\lambda_0}{d} \approx 3.4 $ for which the spontaneous photon is mainly emitted towards the fundamental NFGM, $\text{HE}_{11}$ (see Fig. \ref{FigGammaOrtho}d). 
Intensity patterns Figs. \ref{IntensityOrtho1}(a,b) and Figs. \ref{IntensityOrtho1}(c,d) observed in free-space and in the vicinity of the nanofiber, respectively, have the same global structure. The presence of the nanofiber, however, induces a spatial modulation in Figs. \ref{IntensityOrtho1}(c,d) which results from the interference between (i) 
the light emitted by the ring and directly propagating through free-space towards the observation plane and (ii) the light reaching the observation plane after reflection onto the nanofiber. At the chosen distances $y=1.5 \rho$ and $y=5\rho$ the guided modes are too weak to play a significant role. By contrast, the intensity pattern observed in the $xy$-plane tends to focus around the fiber (Figs. \ref{IntensityOrtho1} g-i) compared to the free-space case (Figs. \ref{IntensityOrtho1} e-f). In particular, far from the ring, this pattern is strongly dominated by the fundamental NFGM HE$_{11}$ as clearly seen in Fig. \ref{IntensityOrtho1}i. 

As already mentioned at the beginning of this section, nanorings of higher $N$'s (up to $N=10$) present the same qualitative behavior as the ring considered above in the collective regime. The main difference is the appearance of new REMs $n\geq3$  which are  strongly subradiant in the collective regime of interest, \emph{i.e.}  $\frac{\lambda_0}{d}\geq2$,  as previously observed in free-space \cite{MPO19}, and therefore do not play any role in the emission of the ring.  In quantum network scenarios, such modes could, however, serve as stable memory states onto which guided photonic excitations could be mapped after being collected by REMs strongly coupled to NFGMs.

Finally, in the Appendix, we present the results we obtained for REM $n=0$ as well as for longitudinal atomic dipoles (Fig. \ref{ConfigSystem}b).

\section{Two rings} \label{SecTwoRings}
\begin{figure*}
    \includegraphics[width = 1\textwidth]{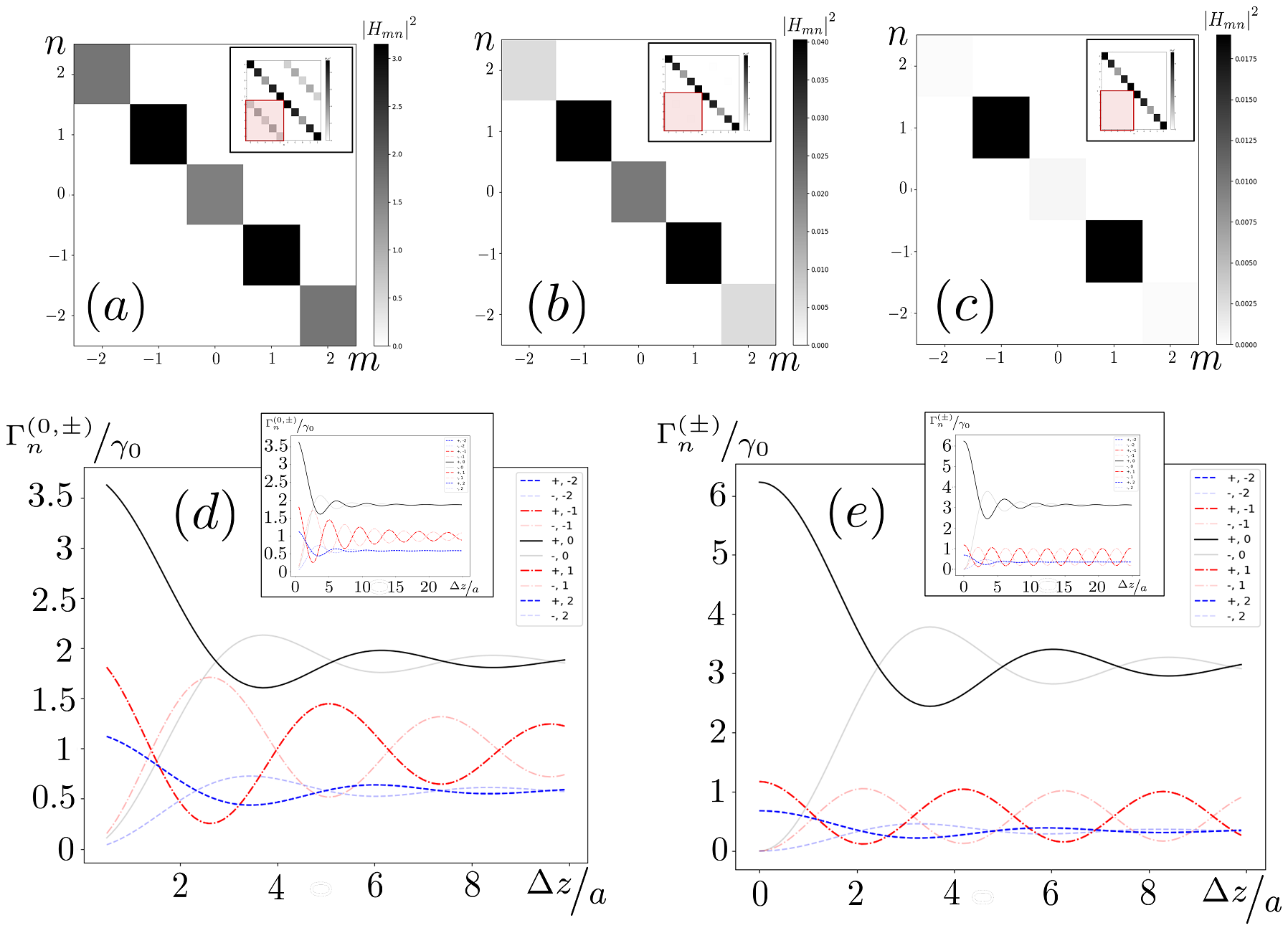}
    \caption{Coupling of two identical nanorings of $N=5$ two-level atoms with orthoradial dipoles. Lower left $N\times N$ block in the effective $2N\times2N$ Hamiltonian expressed in the single-ring eigenmode basis for $\lambda_0 = 3.4\; d$ and different values of the lateral distance between the rings : (a) $\Delta z = a$, (b) $\Delta z = 10 a$, (c) $\Delta z = 20 a$. Inserts show the full effective Hamiltonian, $H_{\text{eff}}$. Terms of  $H_{\text{eff}}$ are plotted in arbitrary units. Decay rates of the two-ring REMs 
$\left(\pm,n=0,1,2\right)$ in free-space,   $\Gamma_{n}^{\left(0,\pm\right)}$ (d), and in the presence of the ONF,  $\Gamma_{n}^{\left(\pm\right)}$ (e), over the single-atom spontaneous emission rate in free space, $\gamma_0$, are plotted as functions of 
$\frac{\Delta z}{a}$ 
for $\lambda_0 = 3.4 \;d$. Inserts show the same functions plotted on a larger range of $\frac{\Delta z}{a}$.}    
    \label{2RingsOrthoHE11}
\end{figure*}

We now consider the case of two identical rings $(I,II)$  positioned at  the lateral distance $\Delta z$  from each other (Fig. \ref{ConfigSystem}c).  The atoms of the lower and upper rings are labeled from $1$ to $N$ and from $\left(N+1\right)$ to $2N$, respectively.
Assuming the two-ring system is initially singly-excited, it evolves according to the same effective Hamiltonian, $H_\text{eff}$, as in the previous case, with the difference that the sums over $i$ and $j$ in Eq.(\ref{Heffectif}) now run from $1$ to $2N$. Defining the REMs of the first and second rings 
as 
$\left|\Psi_{n}^{\left(I\right)}\right\rangle   \equiv \frac{1}{\sqrt{N}}\sum_{j=1}^{N}e^{\text{i} \frac{2 \pi n (j-1)}{N}}\left|e_{j}\right\rangle$ and $\left|\Psi_{n}^{\left(II\right)}\right\rangle  \equiv \frac{1}{\sqrt{N}}\sum_{j=1}^{N}e^{\text{i} \frac{2\pi n (j-1)}{N} }\left|e_{j+N}\right\rangle $,
one 
shows that 
\[
H_\text{eff}\left|\Psi_{n}^{\left(I,II\right)}\right\rangle = -\frac{3 \pi \hbar \gamma_0}{k_0} \left[ \lambda_n\left|\Psi_{n}^{\left(I,II\right)}\right\rangle+  \nu_n\left|\Psi_{n}^{\left(II,I\right)}\right\rangle \right]
\]
with $\lambda_n$'s given in Eq. (\ref{eqlambda})  and 
\begin{equation}
\nu_{n}=\frac{1}{N}\sum_{k,l=1}^{N}g_{k+N,l}e^{-\text{i} \frac{2 \pi n (k-l) }{N}} \label{Eqnu}
\end{equation} 
In the 
basis $\left\{ \left| \Psi^{(I)}_{n= \lceil -\frac{N-1}{2}\rceil \cdots \lceil \frac{N-1}{2}\rceil } \right\rangle \right\}$ $ \cup $  $\left\{ \left| \Psi^{(II)}_{n= \lceil -\frac{N-1}{2}\rceil \cdots \lceil \frac{N-1}{2}\rceil} \right\rangle \right\} $,  $H_\text{eff}$  hence splits into four diagonal blocks (see Figs. \ref{2RingsOrthoHE11} a-c) : the two blocks on the diagonal are identical and contain the $\lambda_n$'s, the off-diagonal blocks are identical and contain the $\nu_n$'s.
An excitation  initially stored in a given REM $n$ of one ring is therefore  mapped into  the  same REM $n$ of the other ring, and back and forth until it is lost.
Contrary to the $\lambda_n$'s, the coupling terms $\nu_n$'s depend on the ring-ring separation, $\Delta z$. At short distance (Fig. \ref{2RingsOrthoHE11}a), $\nu_n$'s include free-space, radiation, and guided mode contributions. As $\Delta z$ increases (Fig. \ref{2RingsOrthoHE11}b), the free-space and radiation mode components progressively vanish and  the $\left\|\nu_n \right\|$'s decrease.
Finally, at large separation $\Delta z$, only NFGMs contribute significantly. In the case shown in Fig. \ref{2RingsOrthoHE11}  -- orthoradial atomic dipoles and $\lambda_0 = 3.4 d$ -- since only  $\text{HE}_{11}$ exists which couples to REMs $n=\pm 1$,  only the terms $\nu_{n=\pm1}$ remain (Fig. \ref{2RingsOrthoHE11}c). Similar observations can be made in other cases, e.g. at $\lambda_0 = 1.8 d$  for which the ONF supports more guided modes, or with dipoles oriented along the ONF axis (see Appendix).

The two-ring-system REMs  are formed as the  symmetric and antisymmetric superpositions of single-ring REMs of same $n$,  \emph{i.e.}  $\left|\Psi_{n}^{\left(\pm\right)}\right\rangle\equiv \frac{1}{\sqrt{2}} \left( \left|\Psi_{n}^{\left(I\right)}\right\rangle \pm \left|\Psi_{n}^{\left(II\right)}\right\rangle \right)$,  for $n=\lceil - \frac{N-1}{2} \rceil \cdots \lceil  \frac{N-1}{2} \rceil$.  The associated eigenvalues are   $-\frac{3\pi \hbar \gamma_0}{k_0} \mu_n^{(\pm)}$  with
\[
\mu_{n}^{\left(\pm\right)}\equiv \left(\lambda_{n}\pm\nu_{n}\right)=\frac{1}{N}\sum_{k,l=1}^{N}e^{-\text{i}n\left(\theta_{k}-\theta_{l}\right)}\left(g_{kl}\pm g_{k+N,l}\right)
\]
and the decay rates are   $\Gamma_n^{\left( \pm \right)} \equiv - \frac{6 \pi \hbar \gamma_0}{k_0} \text{Im}\mu_n ^{\left( \pm \right)}$.
In Fig. \ref{2RingsOrthoHE11}, we plot the decay rates of the REMs  $\left|\Psi_{n}^{\left(\pm\right)}\right\rangle$  in free space, $\Gamma_n^{\left(0, \pm \right)}$ 
(d), and in the presence of the nanofiber, $\Gamma_n^{\left(\pm \right)}$ (e), over the free-space single-atom spontaneous emission rate, $\gamma_0$, as functions of  $\frac{\Delta z}{a}$  for  $\lambda_0 = 3.4d
$   (inserts show the variations 
over a wider range of $\Delta z$).
In the (unphysical) limit $\Delta z\rightarrow 0$, the two rings merge into a single one and $\nu_n\rightarrow\lambda_n$ for all $n$'s, whence $\Gamma_n^{(-)}\rightarrow 0$   and  $\Gamma_n^{(+)}\rightarrow 2 \Gamma_n$.
As $\Delta z$ increases,  decay rates oscillate around their single-ring value, $\Gamma_n^{\left( 0 \right)}$ in free space (d) and $\Gamma_n^{\left( t \right)}$ in the vicinity of the fiber (e). These oscillations vanish at large separation in free-space (insert of d) as well as in the vicinity of the fiber (insert of e)  except for the REMs $(n=1, \pm)$.  Indeed, as mentioned above, the oscillating parts in $\Gamma_n^{\left( \pm \right)}$'s are the coupling terms $\nu_n$'s. The latter vanish when $\Delta z\gg a$ for those REMs which couple only to free-space (Fig. \ref{2RingsOrthoHE11}d)  and radiation modes (REMs $n=0,2$  in Fig. \ref{2RingsOrthoHE11}e). By contrast, they remain nonzero for REMs which couple to NFGMs, \emph{i.e.}  here REMs $n=1$ coupling to $\text{HE}_{11}$ (Fig. \ref{2RingsOrthoHE11}e). Moreover, for free-space and radiation mode contributions, the spatial period of these oscillations is the wavelength $\lambda_0$ while for NFGM contributions it coincides with  $\frac{2\pi}{\beta}$, $\beta$  being  the propagation constant which satisfies the dispersion relation (Fig. \ref{FigGuidedModes}). 
In the case considered in Fig. \ref{2RingsOrthoHE11}e, REMs $\left( n=0,2 ; \pm \right)$  couple only to radiation and free-space modes and their decays exhibit damped oscillations with the period $\lambda_0$, while REMs $\left( n=1 ; \pm \right)$ couple predominantly to NFGMs HE$_{11}$  and show persistent oscillations with the period $\frac{2 \pi}{\beta_{\text{HE}_{11}}} \approx 0.95 \lambda_0$. Note that, though not visible in Fig. \ref{2RingsOrthoHE11}, beatings can be observed between free-space/radiation and guided mode contributions at moderate separation $\Delta z$  (see Appendix).

Finally,  we observe in Fig.  \ref{2RingsOrthoHE11}e that at $\frac{\Delta z}{a}=\frac{2\pi m}{\beta a}\approx  4.2 m$ for $m\in \mathbb{Z}$, REMs $\left(n=1;\pm \right)$ are slightly super- $\left( \Gamma_\text{sup} \approx 1.1 \gamma_0\right )$  and subradiant $\left( \Gamma_\text{sub}\approx0.2\gamma_0 \right)$, respectively, and the opposite at $\frac{\Delta z}{a}= \frac{2\pi}{\beta a} \left( m+\frac{1}{2} \right)$. In addition, we note that one can easily switch from $\left| \Psi^{\left(\pm\right)}_{n=1} \right\rangle$ to $\left| \Psi^{\left(\mp\right)}_{n=1} \right\rangle$ by imposing a $\pi$ phase shift on the excited state of one ring. Therefore, by adjusting the ring separation $\Delta z$,  one can 
map an excitation from the NFGM $\text{HE}_{11}$  into the superradiant branch of REM $n=1$ before transfering it into the subradiant state via the application of a laser beam on one of the rings,  
hence protecting it from losses towards radiation modes for about $\tau\approx5\gamma_0^{-1}$.  The excitation lifetime enhancement we obtain here is indeed moderate. Our findings, however, suggest that larger -- though finite -- arrays of nanorings should give access to a high level of protection while offering switchable, selective, and reliable communication channels via guided modes.

\section{Conclusion \label{SecConclusion}}

The collective emission of one and two nanorings comprising of $N=5$  two-level atoms surrounding a silica ONF was theoretically investigated. The radiation eigenmodes of a single ring were shown to couple only to specific guided modes according to their symmetry, while the emission towards radiation modes can be greatly suppressed by properly tuning the parameters of the system. Two rings can couple even at a large separation distance via guided modes, and this results in an enhanced sub/superradiant character of the radiation eigenmodes with respect to the single-ring case. We believe that the combination of the mode selectivity offered by the single-ring configuration with the guided mode-assisted long-range collective coupling of arrays of such rings is the optimal combination. 
holds great potential for efficient and versatile quantum photonic memories and networks.  While an experimental realization of the proposed setup is challenging, it is worth investigating using either a system of optical tweezers with specific polarizations to create five traps around the ONF or, alternatively, exploring higher order mode interference that could lead to such a configuration by emptying all other traps along the length of the nanofiber via an optical pushing beam.

Other configurations including more atomic levels and
 more rings with tilted angular distributions will be considered in the future for they may give rise to interesting effects, including enhanced chiral coupling of guided light with the atomic system. The presumably highly mode-selective reflection and transmission properties of a stack of nanorings will also  be addressed. 
 Finally the multi-excitation regime and the potential use of the nanoring configuration for the implementation of two-photon gates shall 
be investigated.

\appendix

\section{Supplementary numerical results}
In this Appendix we briefly present complementary numerical results we obtained  in cases not considered in the main text, \emph{i.e.}   orthoradial dipoles at $\lambda_0\approx 1.8 d$ (Sec. \ref{SecOrthoradialbis}) and dipoles along the ONF axis (Sec. \ref{SecLongitudinal}).

\subsection{Orthoradial dipoles, $\lambda_0= 1.8 d$}\label{SecOrthoradialbis}
\begin{figure*}
    \includegraphics[width = 1\textwidth]{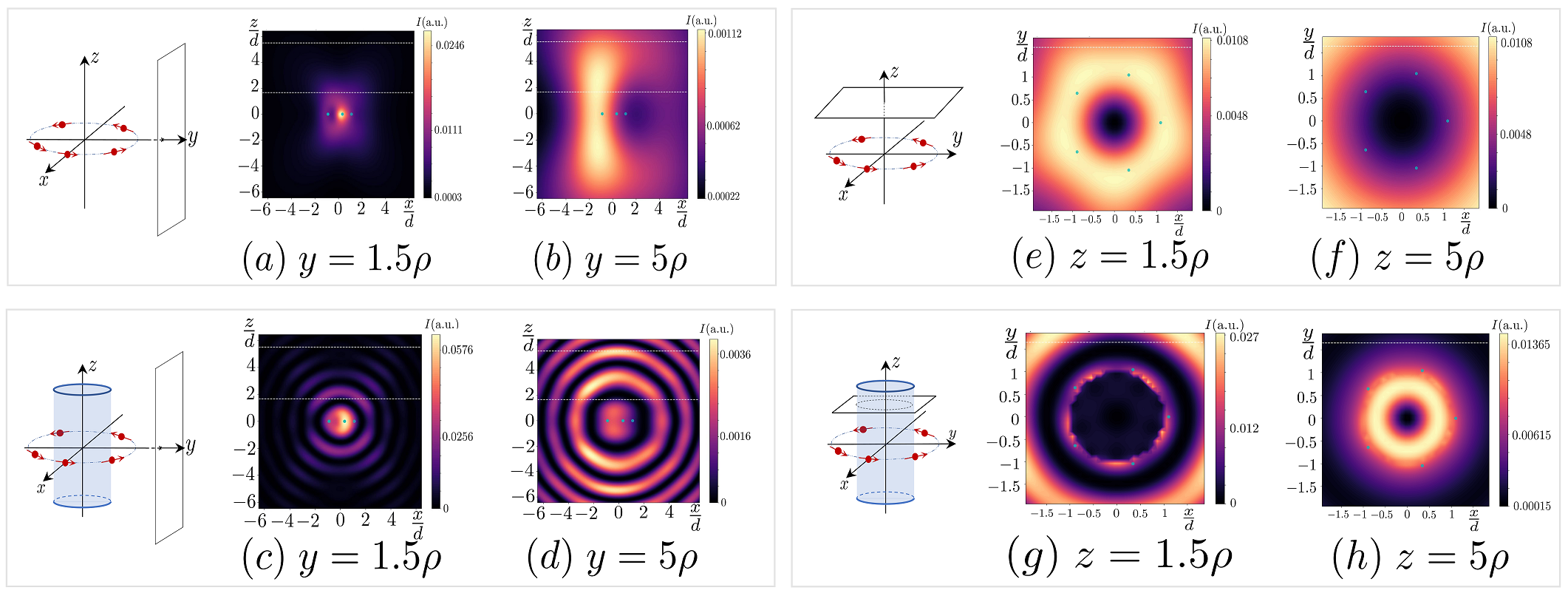}
    \caption{Nanoring of $N=5$ identical two-level atoms with orthoradial dipoles. Radiation patterns of the nanoring prepared in the REM $n=0$ in free space (a,b,e,f) and in the vicinity of an ONF (c,d,g,h). The 2D patterns are represented in the $xz$-plane at $y = 1.5 \rho$ (a,c) and $y = 5 \rho$ (b,d), and in the $xy$-plane at $z = 1.5 \rho$ (e,g) and $z=5 \rho$ (f,h). 
    Intensity is expressed in arbitrary units.   
    In our calculations, we set $\rho=1.1\times a$ whence $d\approx 1.29\times a$, we choose $\frac{\lambda_0}{d} \approx 1.8 $ which corresponds to the maximal ratio $\frac{\Gamma_{n=0}^{\left(\text{TE}_{01}\right)}}{\Gamma^{\left(t\right)}_{n=0}}$ (Fig. \ref{FigGammaOrtho}e) and we take into account 15 radiation modes and all the  NFGMs existing at the considered frequency.}
    \label{IntensityOrtho2}
\end{figure*}

We first consider a single ring of $5$ atoms with  orthoradial dipoles 
and transition wavelength 
$\lambda_0=1.8d$. In that case, the highest relative coupling 
to NFGMs is observed 
for the REM $n=0$ which 
emits  a photon into the mode 
$\text{TE}_{01}$  with a probability $\approx 0.9$  (see  Fig. \ref{FigGammaOrtho}e). In Fig. \ref{IntensityOrtho2}  we plot the corresponding radiation patterns in free-space (a,b,e,f) and close to the ONF (c,d,g,h), in $xz$-plane at $y=1.5 \rho$ (a,c) and $y=5\rho$ (b,d), and in $xy$-plane at $z=1.5\rho$ (e,g) and $z=5\rho$ (f,h).  
As in the case examined in the main text, 
the presence of the ONF 
induces interference fringes  (Figs. \ref{IntensityOrtho2}c,d) not observable in  free-space 
(Figs. \ref{IntensityOrtho2}a,b).  
The intensity pattern observed in the $xy$-plane tends to concentrate around the fiber (Figs. \ref{IntensityOrtho2} g,h) compared to the free-space case (Figs. \ref{IntensityOrtho2} e-f). Far from the ring, this pattern is strongly dominated by the donut-like NFGM TE$_{01}$ (Fig. \ref{IntensityOrtho2}h).

\begin{figure*}
    \includegraphics[width = 1\textwidth]{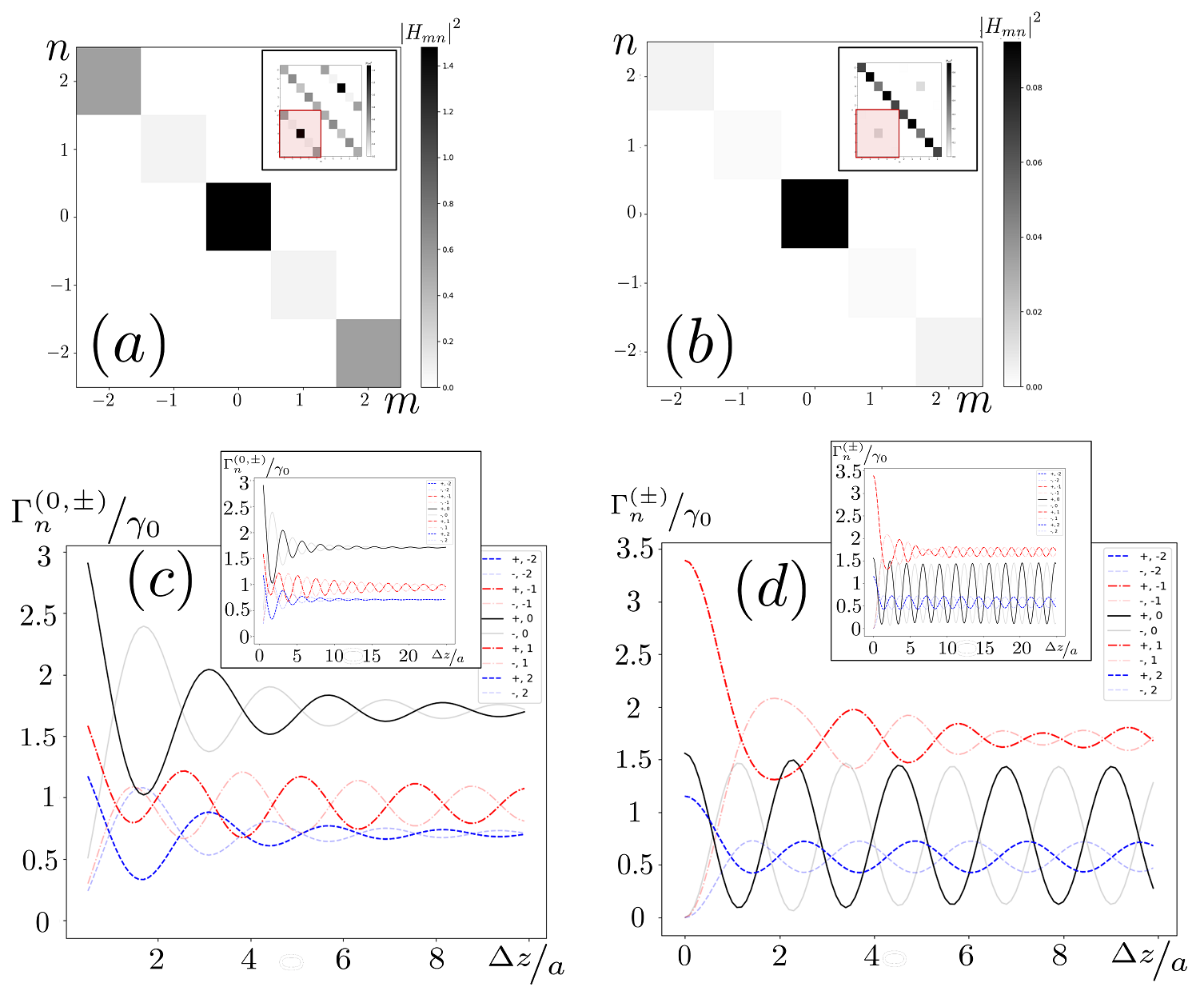}
    \caption{Coupling of two identical nanorings of $N=5$ two-level atoms with orthoradial dipoles. Lower left $N\times N$ block in the effective $2N\times2N$ Hamiltonian expressed in the single-ring eigenmode basis for $\lambda_0 = 1.8\; d$ and different values of the lateral distance between the rings : (a) $\Delta z = a$, (b) $\Delta z = 10 a$. Inserts show the full effective Hamiltonian, $H_{\text{eff}}$. Terms of  $H_{\text{eff}}$ are plotted in arbitrary units. Decay rates of the two-ring REMs  $\left(\pm,n=0,1,2\right)$  in free-space,   $\Gamma_{n}^{\left(0,\pm\right)}$ (c), and in the presence of the ONF,  $\Gamma_{n}^{\left(\pm\right)}$ (d), over the single-atom spontaneous emission rate in free space, $\gamma_0$, are plotted as functions of  $\frac{\Delta z}{a}$  for $\lambda_0 = 1.8 \;d$. Inserts show the same functions plotted on a larger range of $\frac{\Delta z}{a}$.} 
    \label{2RingsOrthoTE01}
\end{figure*}

We now address the two-ring configuration. In Figs. \ref{2RingsOrthoTE01}a,b  we plot  the lower left $N\times N$ block in the effective $2N\times2N$  Hamiltonian which contains the coupling terms $\nu_n$'s on the diagonal.
At large separation $\Delta z=10 a$ (Fig. \ref{2RingsOrthoTE01}b), $\nu_n$'s  contain only  NFGMs contributions, here 
mainly 
due to  
$\text{TE}_{01}$ which couples single-ring REMs $n=0$ to each other, 
 therefore the term $\nu_0$ dominates. But, contrary to the case examined in the main text, other NFGMs also exist, \emph{i.e.} $\text{HE}_{11}$  and   $\text{HE}_{21}$,  which couple single-ring REMs $n=\pm1$ and $n=\pm2$, respectively, and  
$\nu_{\pm1,\pm2}$ therefore do no completely vanish here.  As a consequence, the spatial modulation of the two-ring system decay rates,  $\Gamma_n^{\left(\pm\right)}$,  is observed for all REMs, $\left| \Psi_n^{\left( \pm \right)} \right \rangle$, in the presence of the ONF  even at large separation $\Delta z$ (Fig. \ref{2RingsOrthoTE01}d) while it tends to vanish in free-space (Fig. \ref{2RingsOrthoTE01}c). It is also worth noticing that the period of these modulations is the same for all REMs in free-space, \emph{i.e.}  $\lambda_0\approx 1.8d\approx 2.3 a$ (Fig. \ref{2RingsOrthoTE01}c),  while it varies from one REM to another in the presence of the ONF (Fig. \ref{2RingsOrthoTE01}d). Indeed, as underlined in the main text, at large separation $\Delta z$, the couplings $\nu_n$'s are dominated by the NFGM contributions and therefore oscillate at the corresponding period $\frac{2\pi}{\beta}$, $\beta$ being the propagation constant of the relevant NFGM satisfying the dispersion relation (Fig. \ref{FigGuidedModes}). For $\lambda_0 = 1.8d$  we have  $\beta\approx \left(1.05,1.3,1\right)\times k_0$ for NFGMs  TE$_{01}$, HE$_{11}$ and HE$_{21}$, respectively,  which couple to REMs  $n=0,1,2$, respectively. The respective periods of oscillations of the decay rate $\Gamma_{n=0,1,2}^{\left( \pm \right)}$  are therefore expected to be $\frac{2 \pi}{\beta} \approx \left( \frac{1}{1.05}, \frac{1}{1.2}, 1  \right) \times \lambda_0 \approx \left( 2.2, 1.8, 2.3  \right)\times a$, in good agreement with Fig. \ref{2RingsOrthoTE01}d. 
Moreover, for REMs $\left(n=1,\pm\right)$ at short separation $\Delta z \leq 10 a$  
beatings appear between the free-space mode component -- of period $\lambda_0 \approx 2.3a$ -- and the NFGM HE$_{11}$  contribution -- of period $\frac{2\pi}{\beta}\approx 1.8a$. These beatings are observable because  (i) the period of oscillations differ significantly, (ii) the free-space and guided mode components remain comparable on a sufficiently wide range before unguided modes vanish. By contrast, such beatings cannot be detected with REMs $n=0,\pm2$. 

\subsection{Dipoles along the fiber axis}\label{SecLongitudinal}

\begin{figure*}
    \includegraphics[width=\textwidth]{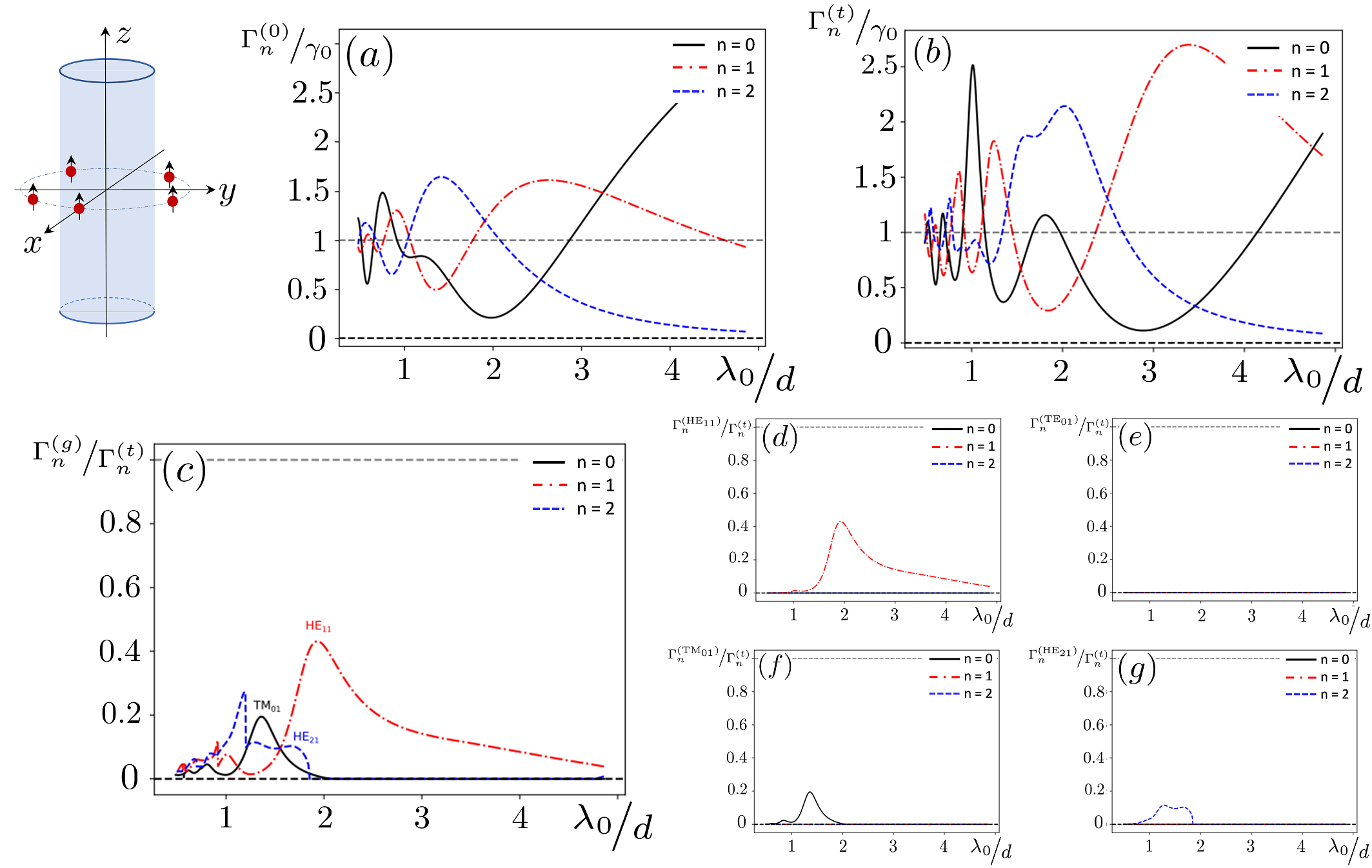}
    \caption{
Nanoring of $N=5$ identical two-level atoms with dipoles oriented along the fiber axis. Total decay rates of the REM $n=0,1,2$ (a) in free space, $\Gamma_{n}^{\left(0\right)}$,  and (b) in the presence of the fiber,  $\Gamma_{n}^{\left(t\right)}$, expressed in units of single-atom spontaneous emission rate in free-space $\gamma_0$.  (c) Ratio of the decay rate of the REM $n=0,1,2$  towards NFGMs,  $\Gamma_{n}^{\left(g\right)}$, over the total decay rate of the same REM, $\Gamma_{n}^{\left(t\right)}$. Ratio of the decay rate of the REM  $n=0,1,2$ towards NFGMs $\lambda$, $\Gamma_{n}^{\left( \lambda \right)}$, over the total emission rate of the same REM, $\Gamma_{n}^{\left(t\right)}$, for $\lambda=\text{HE}_{11} \; (d),\;\text{TE}_{01}\; (e),\;\text{TM}_{01}\; (f),\;\text{HE}_{21}\;(g)$. In our calculations, we set $\rho=1.1\times a$ whence $d\approx 1.29\times a$, we restrict ourselves to the range $\frac{\pi} {5}a \leq\lambda_{0}\leq2\pi a$, and we take into account $15$ radiation modes and all the guided modes supported by the fiber in the considered frequency range. To provide a visual reference, we have plotted the unit value of the different ratios in a dashed grey line in each subfigure.}
    \label{FigGammaLong}
\end{figure*}

\begin{figure*}
    \includegraphics[width = 1\textwidth]{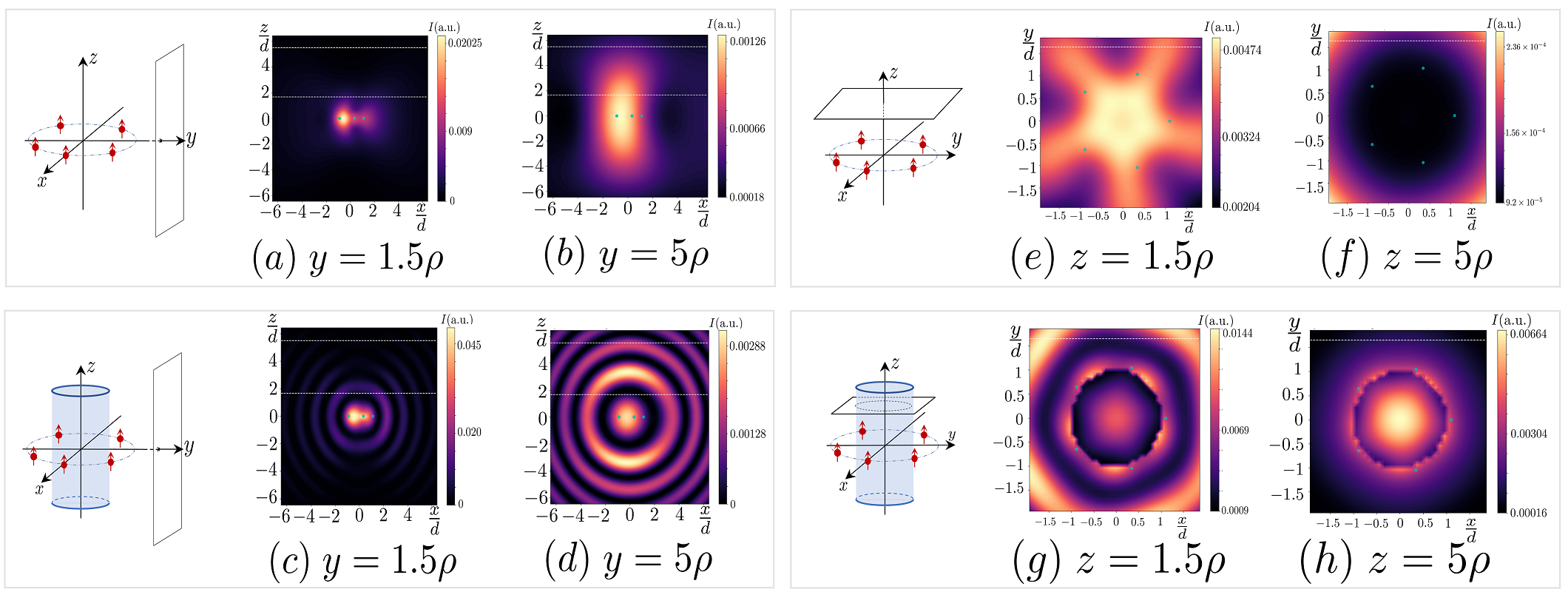}
    \caption{Nanoring of $N=5$ identical two-level atoms with dipoles oriented along the fiber axis. Radiation patterns of the nanoring prepared in the REM $n=1$  in free space (a,b,e,f) and in the vicinity of an ONF (c,d,g,h). The 2D patterns are represented in the $xz$-plane at $y = 1.5 \rho$ (a,c) and $y = 5 \rho$ (b,d), and in the $xy$-plane at $z = 1.5 \rho$ (e,g) and $z=5 \rho$ (f,h).  Intensity is expressed in arbitrary units.  In our calculations, we set $\rho=1.1\times a$ whence $d\approx 1.29\times a$, we choose $\frac{\lambda_0}{d} \approx 1.8 $   which corresponds to the maximal ratio $\frac{\Gamma_{n=1}^{\left(\text{HE}_{11}\right)}}{\Gamma^{\left(t\right)}_{n=1}}$ (Fig. \ref{FigGammaLong}d) and we take into account 15 radiation modes and all the  NFGMs existing at the considered frequency.}
    \label{IntensityLongitudinal1}
    
\end{figure*}

\begin{figure*}
    \includegraphics[width = 1\textwidth]{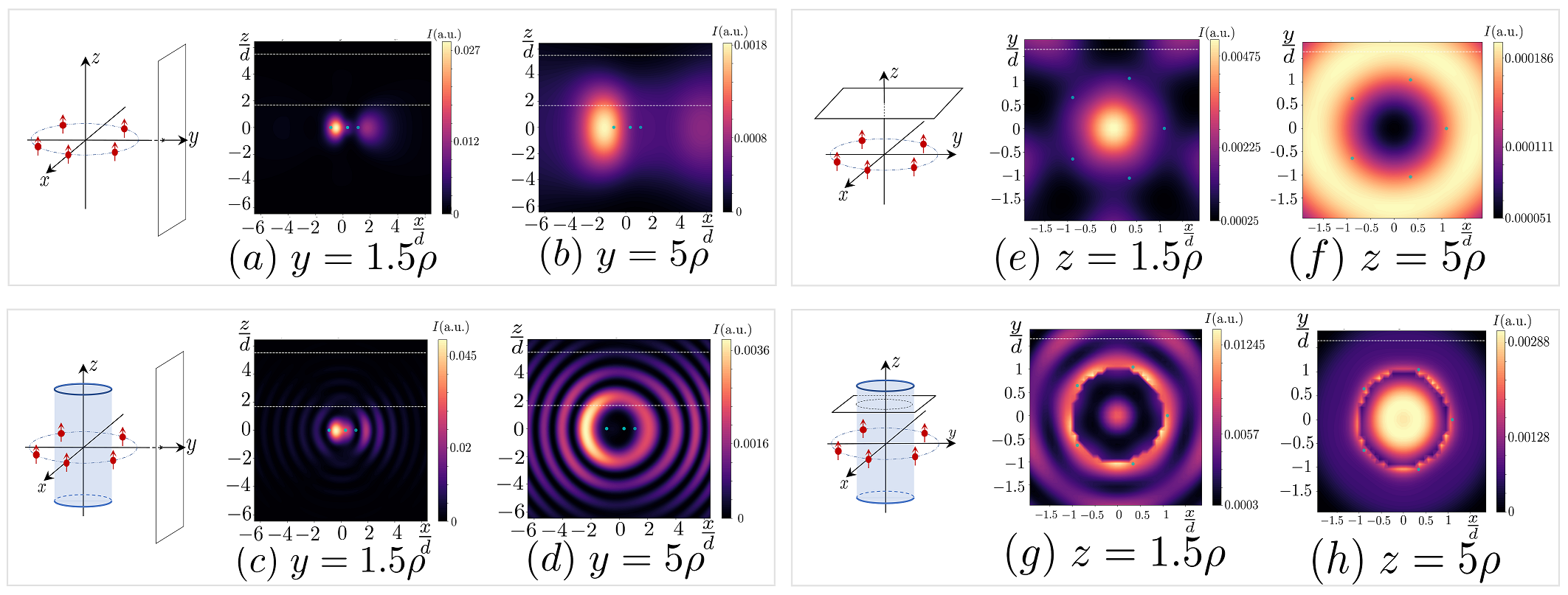}
    \caption{Nanoring of $N=5$ identical two-level atoms with dipoles oriented along the fiber axis. Radiation patterns of the nanoring prepared in the REM $n=0$ in free space (a,b,e,f) and in the vicinity of an ONF (c,d,g,h). The 2D patterns are represented in the $xz$-plane at $y = 1.5 \rho$ (a,c) and $y = 5 \rho$ (b,d), and in the $xy$-plane at $z = 1.5 \rho$ (e,g) and $z=5 \rho$ (f,h).  Intensity is expressed in arbitrary units.  In our calculations, we set $\rho=1.1\times a$ whence $d\approx 1.29\times a$, we choose $\frac{\lambda_0}{d} \approx 1.3 $   which corresponds to the maximal ratio $\frac{\Gamma_{n=0}^{\left(\text{TM}_{01}\right)}}{\Gamma^{\left(t\right)}_{n=0}}$ (Fig. \ref{FigGammaLong}f) and we take into account 15 radiation modes and all the  NFGMs existing at the considered frequency.}
    \label{IntensityLongitudinal2}
    
\end{figure*}

\begin{figure*}
    \includegraphics[width = 1\textwidth]{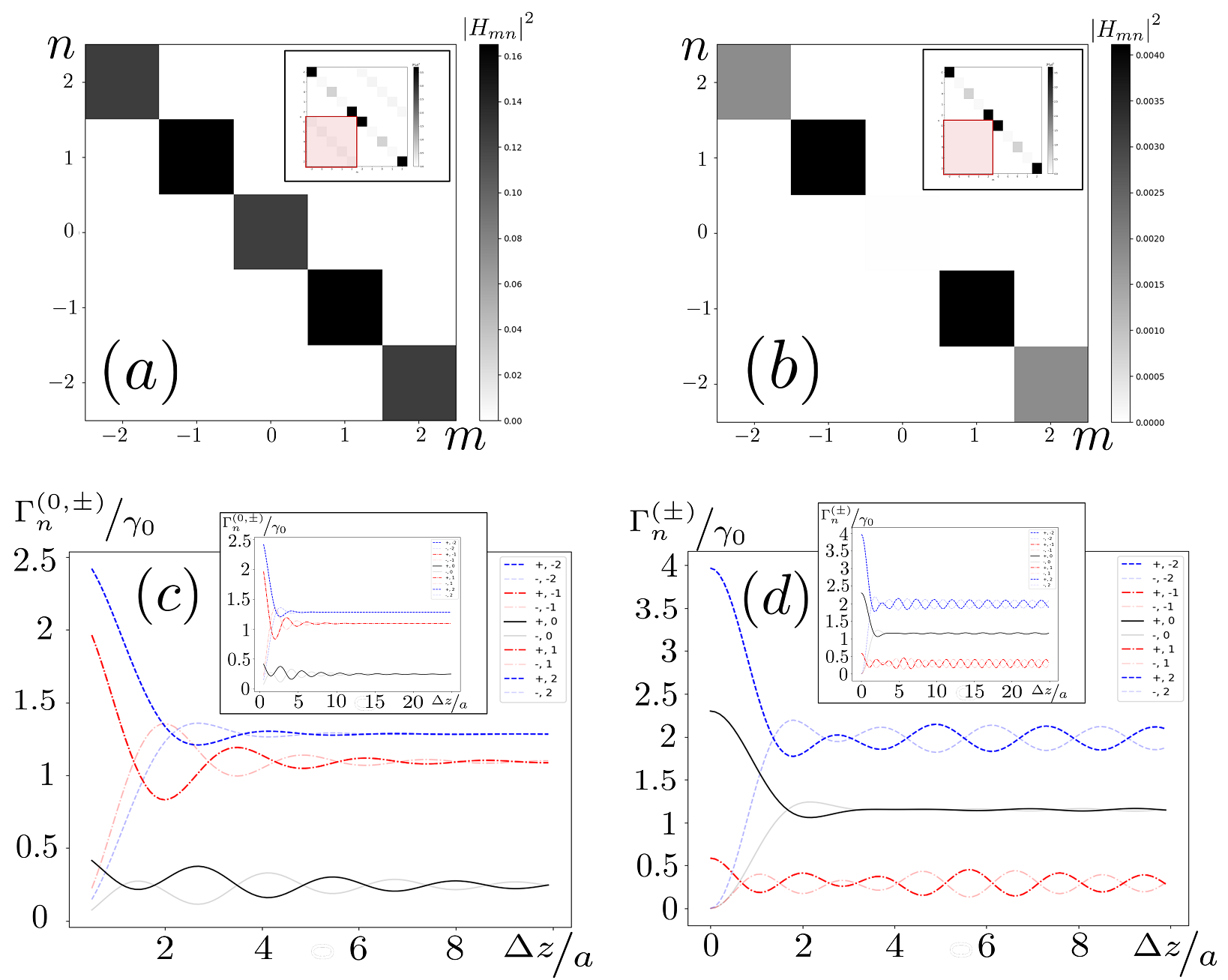}    
    \caption{Coupling of two identical nanorings of $N=5$ two-level atoms with dipoles oriented along the fiber axis. Lower left $N\times N$ block in the effective $2N\times2N$ Hamiltonian expressed in the single-ring eigenmode basis for $\lambda_0 = 1.8\; d$ and different values of the lateral distance between the rings : (a) $\Delta z = a$, (b) $\Delta z = 10 a$. Inserts show the full effective Hamiltonian, $H_{\text{eff}}$. Terms of  $H_{\text{eff}}$ are plotted in arbitrary units. Decay rates of the two-ring REMs  $\left(\pm,n=0,1,2\right)$  in free-space,   $\Gamma_{n}^{\left(0,\pm\right)}$ (c), and in the presence of the ONF,  $\Gamma_{n}^{\left(\pm\right)}$ (d), over the single-atom spontaneous emission rate in free space, $\gamma_0$, are plotted as functions of  $\frac{\Delta z}{a}$  for $\lambda_0 = 1.8 \;d$. Inserts show the same functions plotted on a larger range of $\frac{\Delta z}{a}$.}
 \label{HamTransHE11}
\end{figure*}

\begin{figure*}
    \includegraphics[width = 1\textwidth]{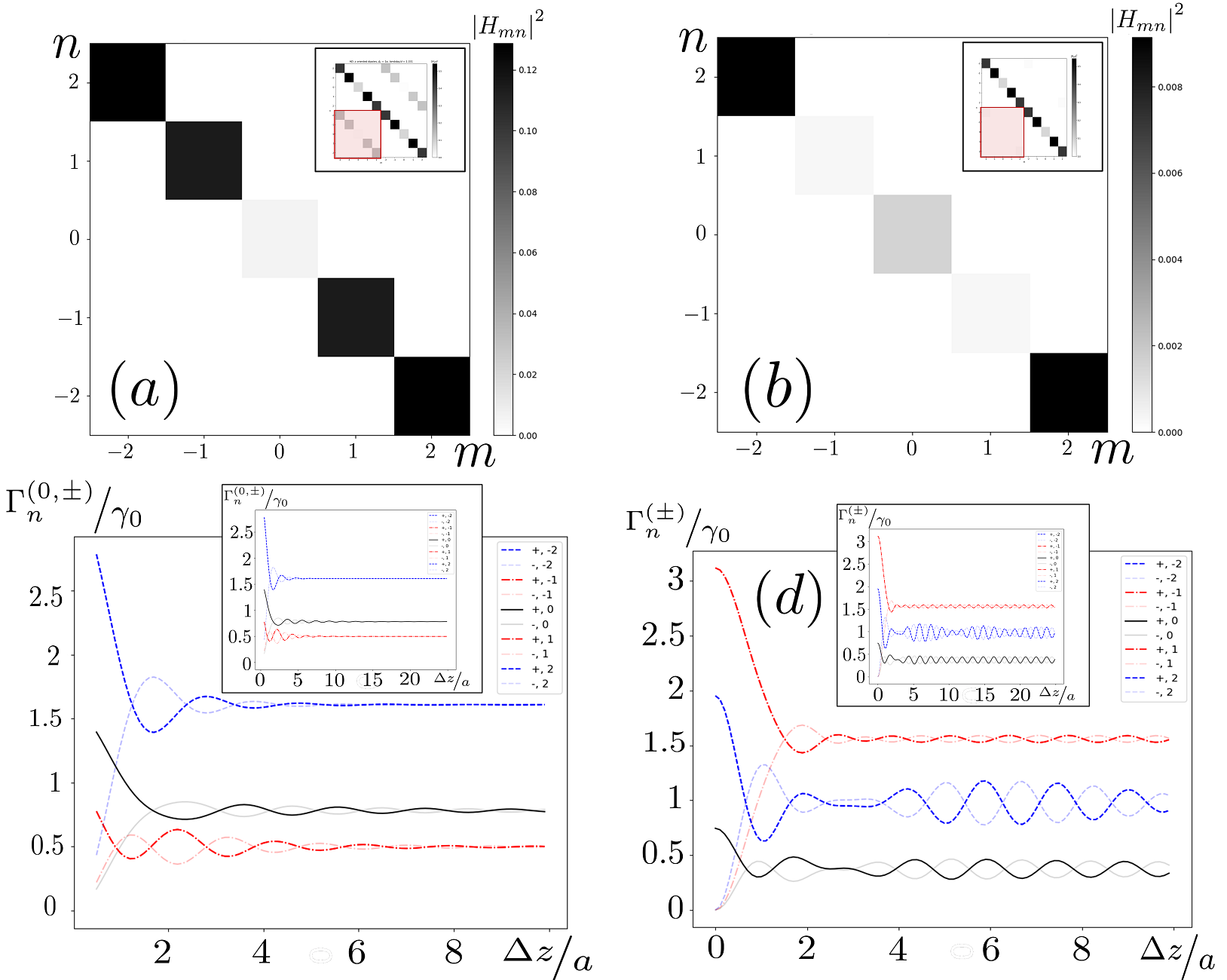}
    \caption{
    Coupling of two identical nanorings of $N=5$ two-level atoms dipoles oriented along the fiber axis. Lower left $N\times N$ block in the effective $2N\times2N$ Hamiltonian expressed in the single-ring eigenmode basis for $\lambda_0 = 1.3\; d$ and different values of the lateral distance between the rings : (a) $\Delta z = a$, (b) $\Delta z = 10 a$. Inserts show the full effective Hamiltonian, $H_{\text{eff}}$. Terms of  $H_{\text{eff}}$ are plotted in arbitrary units. Decay rates of the two-ring REMs  $\left(\pm,n=0,1,2\right)$  in free-space,   $\Gamma_{n}^{\left(0,\pm\right)}$ (c), and in the presence of the ONF,  $\Gamma_{n}^{\left(\pm\right)}$ (d), over the single-atom spontaneous emission rate in free space, $\gamma_0$, are plotted as functions of  $\frac{\Delta z}{a}$  for $\lambda_0 = 1.3 \;d$. Inserts show the same functions plotted on a larger range of $\frac{\Delta z}{a}$.} 
    \label{HamTransTM01}
\end{figure*}

We consider the case of  atomic dipoles oriented along the fiber axis chosen as the $z$-axis. As in the orthoradial orientation case, the  spontaneous emission rates of single-ring REMs $n=0,1,2$  exhibit the same peak structure in free-space (Fig. \ref{FigGammaLong}a) and in the presence of the fiber (Fig. \ref{FigGammaLong}b), though the strongly collective behavior range is shifted towards higher values of the ratio $\frac{\lambda_{0}}{d}$. Moreover, 
the emission features of the atomic ring are enhanced, {\emph i.e.} peaks are higher and minima are deeper.  
 The probability for the atomic ring to emit towards NFGMs, whichever REM it is initially prepared in, is globally weaker than in the orthoradial dipole configuration (Figs \ref{FigGammaLong}c-g). Moreover, REMs $n=0,1,2$ specifically emit towards NFGMs $\text{TM}_{01}$, $\text{HE}_{11}$ and $\text{HE}_{21}$, respectively, as shown in  Fig. \ref{FigGammaLong}d-g. While emission towards the fundamental NFGM $\text{HE}_{11}$ is possible at any frequency, NFGMs $\text{TM}_{01}$ and $\text{HE}_{21}$, however, have cutoff wavelengths, $\frac{\lambda_{0}}{d}\approx2.1$ and $\approx 1.4$, respectively, above which they are not supported by the fiber. The probabilities for the ring prepared in REMs $n=0,1,2$ to emit towards the NFGMs $\text{TM}_{01}$, $\text{HE}_{11}$ and $\text{HE}_{21}$, respectively, reach their maximal values $\approx0.20,0.44,0.10$ around $\frac{\lambda_{0}}{d}\approx1.40,1.90,1.25$, respectively, at which they are subradiant, \emph{i.e.} $\frac{\Gamma_{n=0,1,2}^{\left(t\right)}}{\gamma_{0}}\approx0.40,0.30,0.25$. By contrast, NFGM $\text{TE}_{01}$ does not couple 
 to the atomic ring 
 (Fig. \ref{FigGammaLong}b).
The selective coupling of REMs to specific NFGMs as well as the existence of peaks in Figs. \ref{FigGammaLong}c-g are explained by the same kind of arguments as provided in Sec. \ref{SecSingleRing}. In particular, since atomic dipoles now point along the $z$-axis, they fully couple to NFGM $\text{TM}_{01}$, while they are orthogonal to NFGM $\text{TE}_{01}$.

The radiation patterns of the single-ring REMs $n=1,0$ are plotted in Figs. \ref{IntensityLongitudinal1} and \ref{IntensityLongitudinal2} at wavelengths $\frac{\lambda_0}{d}\approx 1.8, 1.3$ at which they strongly couple to NFGMs HE$_{11}$ and TM$_{01}$, respectively. These plots present the same features as in the orthoradial dipole configuration. Afar from the ONF, the patterns observed in $xz$-plane have the same shape as in free-space though modulated by the interference between the directly transmitted and reflected components of the light emitted by the ring. The patterns observed in $x,y$-plane show the persistence of light concentrated around the fiber axis even at long distance from the ring, by contrast to the free-space case, manifesting the contribution of NFGMs HE$_{11}$ and TM$_{01}$ in the spontaneous emission of REMs $n=1,0$, respectively.

We address the two-ring configuration in Figs. \ref{HamTransHE11}, \ref{HamTransTM01}.  At  $\lambda_0=1.8d$  (resp.  $\lambda_0=1.3d $),  HE$_{11}$ and HE$_{21}$ (resp. TM$_{01}$ and HE$_{21}$) are the NFGMs with highest contributions to ring-ring interaction which couple the REMs $n=\pm1$ and $n=\pm 2$  (resp. $n=0$ and $n=\pm2$)  of one ring to the same REMs of the other ring. As a consequence, the 
terms $\nu_{\pm1,\pm2}$  (resp. $\nu_{0,\pm2}$) dominate at large separation $\Delta z$, see Figs \ref{HamTransHE11}a,b (resp. Figs. \ref{HamTransTM01}a,b).
As in the orthoradial case, decay rates,  $\Gamma_n^{\left(\pm\right)}$,  oscillate with $\Delta z$  in the presence of the ONF, even at large separation  
(Figs. \ref{HamTransHE11}d,\ref{HamTransTM01}d), while these oscillations disappear for increasing $\Delta z$ in free-space (Figs. \ref{HamTransHE11}c,\ref{HamTransTM01}c). These modulations have the same period, $\lambda_0$, for all REMs in free-space. By contrast,  in the presence of the ONF, each REM has its own period, $\frac{2\pi}{\beta}$, imposed by the NFGM contribution. Furthermore, in Figs. \ref{HamTransHE11}d  and \ref{HamTransTM01}d we observe beatings between NFGM contributions and free-space modes at short distance, typically $\Delta z\leq 8 a$ -- out of that range free-space mode contribution to couplings $\nu_n$'s becomes negligible for all REMs. In Fig. \ref{HamTransTM01}d, we observe beatings for REMs $\left(n=2,\pm\right)$ outside the range where free-space modes substantially oscillate. These beatings are indeed due to the interference of NFGMs HE$_{21}$ with radiation modes, particularly the radiation mode with azimuthal number $\nu=\pm3$ , as we could check numerically, the contribution of which remains comparable with the guided mode contribution on a wider range.

\begin{acknowledgments}
This research was funded in part by l’Agence Nationale de la
Recherche (ANR), Project ANR-22-CE47-0011. For the purpose of open
access, the authors have applied a CC-BY public copyright licence
to any Author Accepted Manuscript (AAM) version arising from this
submission. EB and JJM thank Alban Urvoy (Laboratoire Kastler Brossel, Paris, France)  and Klaus M{\o}lmer (University of Copenhagen, Denmark) for enlightening discussions. SNC acknowledges support from OIST Graduate University and  Alexey Vylegzhanin (OIST Graduate University, Japan) and Dylan J. Brown (Imperial College London, UK) for useful discussions.
\end{acknowledgments}


\begin{thebibliography}{10}

\bibitem{LBV24} W. Li, D. Brown, A. Vylegzhanin, Z. Shahrabifarahani, A. Raj, J. Du and S. Nic Chormaic, J. Phys. Photonics \textbf{6} , 021002 (2024).  
\bibitem{FPT12} M. C. Frawley, A. Petcu-Colan, V. G. Truong and S. Nic Chormaic, Opt. Commun. \textbf{285}, 4648–54  (2012).
\bibitem{HFB15} J. E. Hoffman, F. K.  Fatemi, G. Beadie, S. L. Rolston and L. A. Orozco, Optica \textbf{2}, 416 (2015).
\bibitem{LLH04} F. Le Kien, J. Q. Liang, K. Hakuta and V. I. Balykin, Opt. Commun. \textbf{242}, 445 (2004).
\bibitem{LBT17} F. Le Kien, T.  Busch, V. G. Truong and S. Nic Chormaic, Phys. Rev. A \textbf{96}, 023835 (2017). 
\bibitem{NGN16} T. Nieddu, V. Gokhroo and S. Nic Chormaic, J. Opt. \textbf{18}, 053001 (2016).
\bibitem{Kumar_2015} Ravi Kumar, Vandna Gokhroo, Kieran Deasy, Aili Maimaiti, Mary C. Frawley, Ciarán Phelan,  and Síle {Nic Chormaic}, New J. Phys \textbf{17}, 013026 (2015).
\bibitem{YLM12} R. Yalla, F. Le Kien, M. Morinaga and K. Hakuta, Phys. Rev. Lett. \textbf{109}, 063602 (2012).
\bibitem{L14} L. Liebermeister \textit{et al}, Appl. Phys. Lett. \textbf{104}, 031101 (2014).
\bibitem{PSW16} R. N. Patel, T. Schr\"oder, N. Wan, L. Li, S. L. Mouradian, E. H. Chen and D. R. Englund, Light Sci. Appl. \textbf{5}, e16032 (2016).
\bibitem{WVM07} F. Warken, E. Vetsch, D. Meschede, M. Sokolowski and A. Rauschenbeutel, Opt. Express \textbf{15}, 11952 (2007).
\bibitem{PhysRevA.96.043859}
  F. Le Kien, S. S. S. Hejazi, T. Busch, V. G. Truong,  and S. {Nic Chormaic}, Phys. Rev. A \textbf{96}, 043859 (2017). 
\bibitem{LBH04} F. Le Kien, V. I.  Balykin and K. Hakuta, Phys. Rev. A \textbf{70}, 063403 (2004).
\bibitem{FYT07} J. Fu, X. Yin and L. Tong, J. Phys. B: At. Mol. Opt. Phys. \textbf{40}, 4195 (2007).
\bibitem{VRS10} E. Vetsch, D. Reitz, G. Sagué, R. Schmidt, S. T. Dawkins and A. Rauschenbeutel, Phys. Rev. Lett. \textbf{104}, 203603 (2010).
\bibitem{SZL19} E. Stourm, Y. Zhang, M. Lepers, R. Guérout, J. Robert, S. Nic Chormaic, , K. Mølmer, E. Brion, 
J. Phys. B: At. Mol. Opt. Phys. \textbf{52}, 045503 (2019). 
\bibitem{SLR20} E. Stourm, M. Lepers, J. Robert, S. Nic Chormaic, K. Mølmer, E. Brion, Phys. Rev. A \textbf{101}, 052508 (2020).
\bibitem{SLR23} E. Stourm, M. Lepers, J. Robert, S. Nic Chormaic, K. Mølmer, and E. Brion, New J. Phys. \textbf{25}, 023022 (2023).
\bibitem{PhysRevResearch.2.012038} K. S. Rajasree, T. Ray, K. Karlsson, Kristoffer, J. L. Everett and S. {Nic Chormaic}, Phys. Rev. Res. \textbf{2}, 012038 (2020). 
\bibitem{VBR23} A. Vylegzhanin, D. J. Brown, A. Raj, D. F. Kornovan, J. L. Everett, E. Brion, J. Robert, S. Nic Chormaic, Optica Quantum \textbf{1}, 6 (2023). 
\bibitem{SPI23}  A. S. Sheremet, M. I.  Petrov, I. V.  Iorsh, A. V.  Poshakinskiy and A. N. Poddubny, Rev. Mod. Phys. \textbf{95}, 015002 (2023).
\bibitem{AMA17} A. Asenjo-Garcia, M. Moreno-Cardoner, A. Albrecht, H. J. Kimble, and D. E. Chang, Phys. Rev. X \textbf{7}, 031024 (2017).
\bibitem{SBF17} P. Solano, P. Barberis-Blostein, F. K. Fatemi, L. A. Orozco, and S. L. Rolston, Nat. Commun. \textbf{8}, 1857 (2017).
\bibitem{CRC19} N. V. Corzo, J. Raskop, A. Chandra, A. S. Sheremet, B. Gouraud, and J. Laurat, Nature (London) \textbf{566}, 359 (2019).
\bibitem{GMN15} B. Gouraud, D. Maxein, A. Nicolas, O. Morin, and J. Laurat,  Phys. Rev. Lett. \textbf{114}, 180503 (2015).
\bibitem{SCA15} C. Sayrin, C. Clausen, B. Albrecht, P. Schneeweiss, and A.
Rauschenbeutel, Optica \textbf{2}, 353 (2015).
\bibitem{CGC16} N. V. Corzo, B. Gouraud, A. Chandra, A. Goban, A. S. Sheremet, D. V. Kupriyanov, and J. Laurat, Phys. Rev. Lett. \textbf{117}, 133603 (2016).
\bibitem{SML12} A. S. Sheremet, A. D. Manukhova, N. V. Larionov, and D. V. Kupriyanov, Phys. Rev. A \textbf{86}, 043414 (2012).
\bibitem{KSP16} D. F. Kornovan, A. S. Sheremet, and M. I. Petrov,  Phys. Rev. B \textbf{94}, 245416 (2016).
\bibitem{JCC18} H.H. Jen, M. S. Chang,  and Y. C. Chen, Sci. Rep. \textbf{8}, 9570 (2018). 
\bibitem{PSR19} D. Plankensteiner, C. Sommer, M. Reitz, H. Ritsch, and C. Genes,  Phys. Rev. A \textbf{99}, 043843 (2019).
\bibitem{ZM19} Y.-X. Zhang and K. Mølmer, Phys. Rev. Lett. \textbf{122}, 203605 (2019).
\bibitem{FSK21} Y. A. Fofanov, I. M. Sokolov, R. Kaiser, and W. Guerin, Phys. Rev. A \textbf{104}, 023705 (2021).
\bibitem{PCP85} D. Pavolini, A. Crubellier, P. Pillet, L. Cabaret, and S. Liberman, Phys. Rev. Lett. \textbf{54}, 1917 (1985).
\bibitem{DB96} R. G. DeVoe and R. G. Brewer, Phys. Rev. Lett. \textbf{76}, 2049 (1996).
\bibitem{GAK16} W. Guerin, M. O. Araújo, and R. Kaiser, Phys. Rev. Lett. \textbf{116}, 083601 (2016).
\bibitem{RWR20} J. Rui, D. Wei, A. Rubio-Abadal, S. Hollerith, J. Zeiher, D. M. Stamper-Kurn, C. Gross, and I. Bloch, Nature (London) \textbf{583}, 369 (2020).
\bibitem{FGH21} G. Ferioli, A. Glicenstein, L. Henriet, I. Ferrier-Barbut, and A. Browaeys, Phys. Rev. X \textbf{11}, 021031 (2021).
\bibitem{NLO19}J. A. Needham, I. Lesanovsky and B. Olmos, New J. Phys. \textbf{21}, 073061 (2019).
\bibitem{ZM20} Y.-X. Zhang and K. Mølmer, Phys. Rev. Lett. \textbf{125}, 253601 (2020).
\bibitem{KSK21} D. F. Kornovan, R. S. Savelev, Y. Kivshar, and M. I. Petrov, ACS Photonics \textbf{8}, 3627 (2021).
\bibitem{FFB23} N. Fayard, I. Ferrier-Barbut, A. Browaeys, and J.-J. Greffet, Phys. Rev. A \textbf{108}, 023116 (2023).
\bibitem{KPL19} Y. Ke, A. V. Poshakinskiy, C. Lee, Y. S. Kivshar, and A. N. Poddubny,  Phys. Rev. Lett. \textbf{123}, 253601 (2019).
\bibitem{SR16} R. T. Sutherland and F. Robicheaux, Phys. Rev. A \textbf{94}, 013847 (2016).
\bibitem{AHC17} A. Asenjo-Garcia, J. D. Hood, D. E. Chang, and H. J. Kimble, Phys. Rev. A \textbf{95}, 033818 (2017).
\bibitem{KCL19} D. F. Kornovan, N. V. Corzo, J. Laurat, and A. S. Sheremet, Phys. Rev. A \textbf{100}, 063832 (2019).
\bibitem{ZYM20} Y.-X. Zhang, C. Yu, and K. Mølmer, Phys. Rev. Res. \textbf{2}, 013173 (2020).
\bibitem{P20} A. N. Poddubny, Phys. Rev. A \textbf{101}, 043845 (2020).
\bibitem{FJR16} G. Facchinetti, S. D. Jenkins, and J. Ruostekoski, Phys. Rev. Lett. \textbf{117}, 243601 (2016).
\bibitem{BGA15} R. J. Bettles, S. A. Gardiner, and C. S. Adams, Phys. Rev. A \textbf{92}, 063822 (2015).
\bibitem{BR20} K. E. Ballantine and J. Ruostekoski, Phys. Rev. Res. \textbf{2}, 023086 (2020).
\bibitem{MPO19}M. Moreno-Cardoner, D. Plankensteiner, L. Ostermann, D. E. Chang, and H. Ritsch, Phys. Rev. A \textbf{100}, 023806 (2019).
\bibitem{CPM20} J. Cremer, D. Plankensteiner, M. Moreno-Cardoner, L. Ostermann, and H. Ritsch,  New J. Phys. \textbf{22}, 083052 (2020).
\bibitem{HPO20}R. Holzinger, D. Plankensteiner, L. Ostermann, and H. Ritsch, Phys. Rev. Lett. \textbf{124}, 253603 (2020).
\bibitem{MHR22} M. Moreno-Cardoner, R. Holzinger, and H. Ritsch, Opt. Express \textbf{30}, 10779 (2022).
\bibitem{CLO23} M. Cech, I. Lesanovsky, and B. Olmos, Dispersionless subradiant photon storage in one-dimensional emitter chains, Phys. Rev. A \textbf{108}, L051702 (2023).
\bibitem{HPO24} R. Holzinger, J. S. Peter, S. Ostermann, H. Ritsch, and S. Yelin, Opt. Quantum \textbf{2}, 57 (2024).
\bibitem{UKV24} N. Ustimenko, D. Kornovan, I. Volkov, A. Sheremet, R. Savelev, and M. Petrov, Phys. Rev. A \textbf{110}, L011501 (2024).


\bibitem{KDN05} Fam Le Kien, S. Dutta Gupta, K. P. Nayak, and K. Hakuta, Phys. Rev. A \textbf{72}, 063815 (2005).
\bibitem{KR17} Fam Le Kien, A. Rauschenbeutel, Phys. Rev. A \textbf{95}, 023838 (2017).


\bibitem{KKN22}F. Le Kien, D. F. Kornovan, S. Nic Chormaic, and T. Busch,  Phys.
Rev. A \textbf{105}, 042817 (2022).

\bibitem{Tai94}C. T. Tai, Dyadic Green\textquoteright s Functions
in Electromagnetic Theory, 2nd ed. (IEEE Press, Piscataway, NJ, 1994).

\bibitem{KBT17}Fam Le Kien, Thomas Busch, Viet Giang Truong, and Síle
Nic Chormaic,  Phys. Rev. A \textbf{96}, 023835 (2017). 

\end{thebibliography}
\end{document}